\documentstyle[twoside,fleqn,espcrc2,epsf]{article}

\newcommand{\ewxy}[2]{\setlength{\epsfxsize}{#2}\epsfbox[10 30 640 590]{#1}}
\newcommand{\figurebox}[2]{\fbox{\vbox to#2in{\hbox to #1in{\hfil} \vfil}}}

\newcommand{\beq}{\begin{equation}}
\newcommand{\eeq}{\end{equation}}
\newcommand{\beqn}{\begin{eqnarray}}
\newcommand{\eeqn}{\end{eqnarray}}

\def\Vmu{V^{\mu}}
\def\Amu{A^{\mu}}
\def\gq2{g(q^2)} 
\def\pd*{p_{D^*}}

\newcommand{\Vcb}{\mbox{$V_{cb}$}}
\newcommand{\BtoDs}{\mbox{$\bar{B} \to D^*$}}
\newcommand{\BtoDsl}{\mbox{$\bar{B} \to D^* l \bar{\nu}$}}
\newcommand{\BtoDl}{\mbox{$\bar{B} \to D l \bar{\nu}$}}
\newcommand{\kQ}{\mbox{$\kappa_Q$}}
\newcommand{\kQp}{\mbox{$\kappa_{Q'}$}}
\newcommand{\kq}{\mbox{$\kappa_q$}}
\newcommand{\VAcurrent}{\mbox{$\bar{b}(1-\gamma_5) \gamma_{\mu}c$}}
\newcommand{\vdotv}{\mbox{$v \cdot v'$}}
\newcommand{\wisgur}{\mbox{$\xi(\vdotv)$}}
\newcommand{\hA}{\mbox{$h_{A_1}$}}
\newcommand{\hAR}{\mbox{$h_{A_1}^R$}}
\newcommand{\hV}{\mbox{$h_{V}$}}
\newcommand{\hVR}{\mbox{$h_{V}^R$}}
\newcommand{\Om}{\mbox{$O( \frac{1}{m_Q} )$}}
\newcommand{\Omsq}{\mbox{$O( \frac{1}{m_Q^2} )$}}

\newcommand{\rhosq}{\mbox{$\rho^2$}}
\newcommand{\rhosqlat}{\mbox{$\rho^2_{lat}$}}

\newcommand{\wisgurPARAMrho}{\mbox{$\xi_{\rho}$}}
\newcommand{\wisgurPARAMrholat}{\mbox{$\xi_{\rho_{lat}}$}}
\newcommand{\wisgurPARAMrhov}{\mbox{$\xi_{\rho}(\vdotv)$}}
\newcommand{\ZV}{\mbox{$Z_V$}}
\newcommand{\ZA}{\mbox{$Z_A$}}
\newcommand{\ZAinv}{\mbox{$\ZA^{-1}$}}
\newcommand{\ZAinvhAR}{\mbox{$\ZA^{-1} \hAR $}}
\newcommand{\normwisgur}{\mbox{$\xi(1)=1$}}
\newcommand{\capgamma}{\mbox{$\Gamma$}}
\newcommand{\tauB}{\mbox{$\tau_B$}}

\begin{document}

\title{An Extraction of \Vcb\ from the Semi-Leptonic \BtoDs\ Decay}
\author{
UKQCD Collaboration, presented by Nicholas Hazel.
\address{Department of Physics and Astronomy, The University of Edinburgh,
Edinburgh EH9 3JZ, Scotland.}
Edinburgh Preprint 93/538.
}

\begin{abstract}
We present an extraction of \Vcb\ from a lattice calculation of the \BtoDsl\
decay matrix elements. We obtain $|\Vcb| \sqrt{ \frac{ \tauB }{ 1.49 ps } } =
0.037^{+3 \ +5}_{-3 \ -5}$ from a single parameter fit to the new CLEO data.
\vspace{-4mm}
\end{abstract}

\maketitle

\section{Introduction}

The \BtoDsl\ decay proceeds via the spectator process whereby the
bottom quark in the $B$ meson decays, through the weak interaction, to a charm
quark forming a $D$ meson; the common light quark ( $u$ or $d$ ) takes no part
in the interaction. Thus the current matrix element is directly proportional to
the Cabbibo Kobayashi Maskawa $( CKM )$ matrix element \Vcb.

The vector and axial vector current matrix elements can be parametrised in terms
of four form factors, $h_i$'s, which are independent of the meson masses :

\vspace{-2mm}

\beqn
\frac{\langle D^*(v',\epsilon)| \Vmu | \bar{B} (v) \rangle}{\sqrt{m_j m_i}} =
i h_V(v\cdot v') \
\epsilon^{\mu \nu \lambda \sigma}\ \epsilon_{\nu}^{*} v'_{\lambda}
v_{\sigma} & &
\nonumber
\vspace{-2mm}
\eeqn

\vspace{-5mm}

\beqn
\frac{\langle D^*(v',\epsilon)| \Amu | \bar{B} (v) \rangle}{\sqrt{m_j m_i}} =
(1+v\cdot v')h_{A_{1}}(v\cdot v')
\epsilon^{*\mu}  & &
\nonumber
\vspace{-2mm}
\eeqn

\vspace{-5mm}

\beqn
- \epsilon^* \cdot v \{ h_{A_{2}}(v\cdot v')v^{\mu} + 
                        h_{A_{3}}(v\cdot v')v'{}^{\mu} \}, 
\vspace{-2mm}
\eeqn
where $v$ and $v'$ are the meson four-velocities. 
In the limit of infinite heavy quark mass these form factors, and those of the
associated \BtoDl\ decay, are related through an exact spin-flavour symmetry 
to a single universal function \wisgur, the Isgur-Wise function \cite{Wisgur}.

\section{Calculation Details}

We use a $24^3 \times 48$ lattice at $\beta = 6.2$.
This corresponds to an inverse lattice spacing $a^{-1} = 2.73(5) \ GeV$
\cite{UKQCD lat spacing}. This calculation is based on $60\ SU(3)$ gauge
field configurations with quark propagators generated using an
$O(a)$ improved fermion action \cite{action}.

We compute the three point correlators using the standard source method
\cite{source} choosing $t = 24$ as the extension point. We then symmetrize
about this point using Euclidean time reversal. Matrix elements are obtained
by fitting to the form of the three point correlator at large Euclidean time
with meson energy factors constrained to values from two point correlator fits.
We use a local, $O(a)$ improved, weak current of the form \VAcurrent\
and rescale this to the continuum using the renormalization
constants $Z_V$ and $Z_A$.

These preliminary results are for degenerate `bottom'and `charm' quarks,
\kQp\ and $\kQ\ = 0.12900$, and for three values of the light quark mass,
$\kq\ = 0.14144$, $0.14226$ and $0.14262$. We give the $B$ meson lattice momentum 
$(0,0,0)$ and  $(1,0,0)$ and inject
up to $\sqrt{4}$ units at the current operator, the corresponding \vdotv\
lie between $1.0$ and $1.3$.
Statistical errors are calculated using a bootstrap procedure
for 100 samples. Correlations between timeslices are taken into
account.

\section{Axial Vector Form Factor}

In the infinite heavy quark limit the axial vector form factor, \hA, is exactly the
Isgur-Wise function, \wisgur. For finite heavy quark mass there exist two sources
of symmetry breaking; radiative corrections, $R$, from renormalization of the heavy
quark current by hard gluon exchange, and power corrections in the inverse 
heavy quark mass :

\vspace{-2mm}

\beq
\hA = ( \ 1 + R + O( \frac{1}{m_Q} ) \cdot \cdot\ \cdot\ ) \ \wisgur.
\label{corrections}
\eeq

\vspace{-2mm}

The radiative corrections are perturbative QCD corrections and hence
can be evaluated analytically in a model independent way. We calculate these 
using Neubert's short distance expansion of heavy quark currents \cite{Neubert}.
The corrections proportional to inverse powers of the heavy quark mass are
non-perturbative. They are model dependent. However, {\it Luke's theorem } \cite{Luke}
demands that \Om\ corrections to \hA\ vanish at zero recoil leaving corrections
of \Omsq; see \cite{Henning}.

We can define the radiatively corrected axial vector form factor, \hAR, as :

\vspace{-2mm}

\beq
\hAR = \frac{ \hA }{ ( 1 + R ) } = \wisgur.
\eeq

\begin{figure}[tbp]  
\vspace{-8mm}
\ewxy{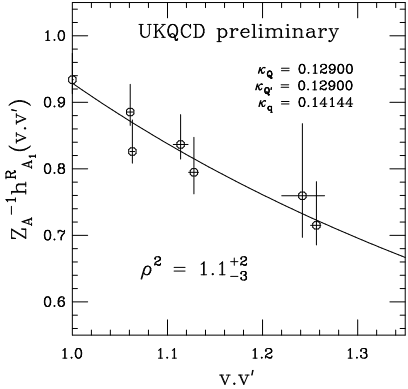}{200pt}
\vspace{-12mm}
\caption{Axial vector form factor}
\vspace{-6mm}
\label{axial form factor}
\end{figure}

We use the BSW \cite{BSW} parametrisation of the Isgur-Wise function :

\vspace{-2mm}

\beq
\wisgurPARAMrhov = \frac{2}{1+\vdotv}\exp{\{(2\rhosq-1)\frac{1-\vdotv}{1+\vdotv} \}}
\eeq
and perform a two parameter fit of the radiatively corrected axial vector form
factor, \ZAinvhAR, to the $BSW$ type model :

%\vspace{-2mm}

\beq
\ZAinvhAR = \ s \ \wisgurPARAMrho,
\label{BSW fit form}
\eeq
where \rhosq\ is the slope parameter at zero recoil and $s =$ \ZAinv; 
using \normwisgur\ from heavy quark symmetry.
Figure \ref{axial form factor} shows a fit to our data for our heaviest light
quark mass.

\section{Axial Vector Current Renormalisation}

\begin{figure}[tb]
\vspace{-8mm}
\ewxy{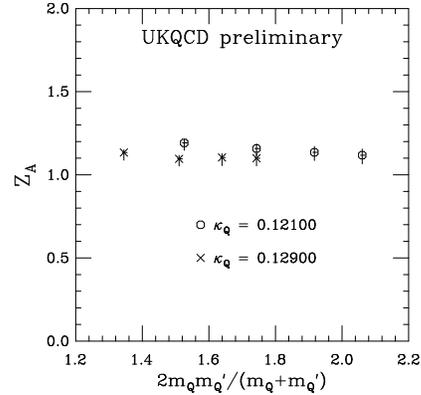}{200pt}
\vspace{-12mm}
\caption{Renormalization constant $Z_A$}
\vspace{-6mm}
\label{renormalization constant} 
\end{figure}

The value of \ZA\ obtained from the above fit will contain \Omsq\ corrections.
Figure \ref{renormalization constant} shows a plot of \ZA\ for a range of
non-degenerate heavy quark kappa values; the `bottom' quark at 
\kQ\ $= 0.12900$ and $0.12100$, four `charm' quark masses and the
light quark at \kq\ $= 0.14144$. The consistency of values suggests that
that \ZA\ is at most only a weak function of the heavy quark masses. This is
particularly important when studying ratios of form factors \cite{Henning}.
\ZA\ appears to have a value of around $1.1$.

\section{Extraction of \Vcb}

The differential decay rate for \BtoDsl, \capgamma, can be expressed in the form :

%\vspace{-2mm}

%\beq
%\frac{ f(\capgamma_{\BtoDsl}) }{ \sqrt{ \tauB } } = \ |\Vcb| \ \wisgur =
%s \ \wisgurPARAMrholat,
\beq
f(\capgamma_{\BtoDsl}) = \ |\Vcb| \ \wisgur = s \ \wisgurPARAMrholat.
\eeq
We extract \Vcb\ from a
one parameter fit, $s$, of the experimental data to our two parameter `$BSW$' model
%(\ref{BSW fit form})% 
constraining \rhosq\ to \rhosqlat, the  value of \rhosq\ obtained from our fit to
\ZAinvhAR.

\begin{figure}[tb]
\vspace{-8mm}
\ewxy{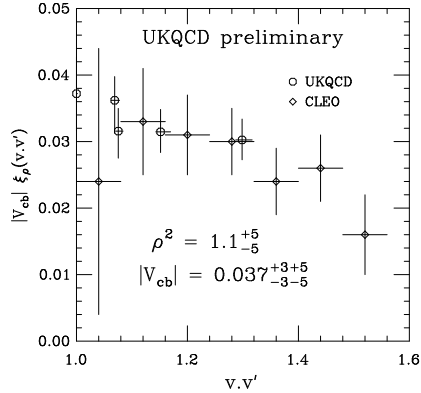}{200pt}
\vspace{-12mm}
\caption{\Vcb\ fit to CLEO data}
\vspace{-6mm}
\label{CLEO Vcb Fit} 
\end{figure}

\begin{figure}[tb]
\vspace{-8mm}
\ewxy{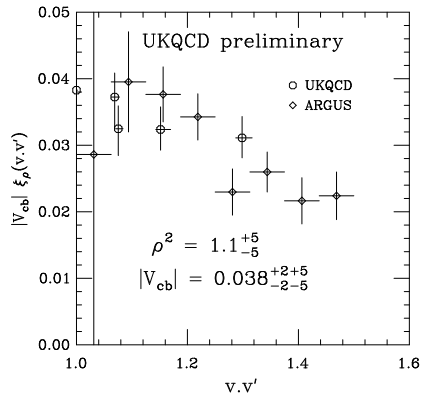}{200pt}
\vspace{-12mm}
\caption{\Vcb\ fit to ARGUS data}
\vspace{-6mm}
\label{ARGUS Vcb Fit} 
\end{figure}

Before extracting \Vcb\ we extrapolate the light quark mass to the
chiral limit and obtain $\rhosqlat\ = 1.1^{+5}_{-5}$. Figures \ref{CLEO Vcb Fit} and
\ref{ARGUS Vcb Fit}
show \Vcb\ fits to the new CLEO \cite{cleo expt data} and new ARGUS 
\cite{argus expt data} data respectively. 

We obtain from the fit to the CLEO data~:

\beq
|\Vcb| \sqrt{ \frac{ \tauB }{ 1.49 ps } } = 0.037^{+3 \ +5}_{-3 \ -5},
\eeq
\nopagebreak
where \tauB\ is the lifetime of the $B$ meson. 
The first set of errors are due to experimental uncertainties while
the second are due to uncertainties in our lattice determination of \rhosq.
The value of \Vcb\ obtained from a fit to the ARGUS data is in excellent 
agreement.

\section{Vector Form Factor}

%Finally Figure \ref{vector form factor} shows a plot of the radiately 
%corrected vector form factor, \hVR\, for our heaviest light quark mass.
%We cannot extract \ZV\ because \hV\ is not protect by Luke's theorem, as was
%\ZA, from large \Om\ corrections. Assuming \ZV\ is close to $1.0$, we 
%observe \Om\ corrections of $30-40~\%$, in line with Neubert's predictions
%\cite{Neubert again}.

\begin{figure}[tb]
\vspace{-8mm}
\ewxy{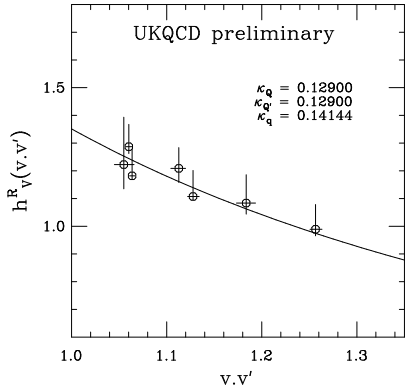}{200pt}     
\vspace{-12mm}
\caption{Vector form factor} 
\vspace{-6mm}
\label{vector form factor}
\end{figure}

Finally Figure \ref{vector form factor} shows a plot of the radiatively
corrected vector form factor, \hVR, for our heaviest light quark mass.
\hV\ is not protected by {\it Luke's theorem } and contains large \Om\ corrections.
Correcting for \ZV\, from our study of the \BtoDl\ decay
matrix elements \cite{wisgur paper}, we observe \Om\ corrections of $30-40~\%$.
This is consistent with Neubert's predictions \cite{Neubert again}.

\section{Conclusions}

We have determined \Vcb\ from a lattice computation of the \BtoDsl\
form factors. This value is in good agreement with other estimates;
in particular with our value obtained, assuming heavy quark
symmetry, from the \BtoDl\ decay \cite{Jim}.

\section*{Acknowledgments}
This work was supported by SERC grant GR/H01069, and performed on a
Meiko i860 Computing Surface supported by SERC grant GR/G32779, Meiko
Limited and the University of Edinburgh.  I thank SERC for
financial support throughout my studentship.

\end{document}